\documentclass[aps,prc,twocolumn,reprint,floatfix,superscriptaddress,showpacs]{revtex4-1}
\usepackage{graphicx}
\usepackage[pdftex,colorlinks,citecolor=blue,bookmarks]{hyperref}
\usepackage{bm}
\usepackage{longtable}
\begin{document}
\title{Configuration mixing description of the nucleus  $^{44}$S}

\author{Tom\'as R. Rodr\'iguez} 
\affiliation{GSI Helmholtzzentrum
  f\"ur Schwerionenforschung, D-64291 Darmstadt, Germany} 
\author{J. Luis Egido} 
\affiliation{Departamento de F\'isica Te\'orica, Universidad
  Aut\'onoma de Madrid, E-28049 Madrid, Spain}
\date{\today} 
\begin{abstract}
We study the structure of the neutron rich $^{44}$S isotope with modern configuration mixing methods based on the Gogny interaction including beyond-mean-field effects. Restoration of particle number and rotational symmetries are taken into account as well as shape mixing in the whole triaxial $(\beta_{2},\gamma)$ plane. We obtain a qualitative agreement between the calculated spectrum and the experimental data reported recently. Very extended collective wave functions  in the $(\beta_2,\gamma)$ plane are found for the lowest states, corresponding to very flat potentials, indicating strong configuration mixing and supporting the weakening of the $N=28$ magic number.  
\end{abstract}
\maketitle
Degradation and/or appearance of shell closures in nuclei beyond the valley of stability is one of the main topics in nuclear structure for the last two decades or more. In particular, the magic number $N=28$ associated to the shell closure of a spherical harmonic oscillator plus spin-orbit single particle potential has been extensively studied in neutron rich nuclei both theoretical and experimentally. For instance, the low excitation energy measured for the first $2^{+}$ state in the $^{42}$Si isotope has determined the removal of such a shell closure in this case~\cite{PRL_99_022503_2007}. Indeed, the nucleus $^{44}$S is in between the spherical doubly-magic $^{48}$Ca and the oblate deformed $^{42}$Si isotopes. Therefore, coexistence of different shapes and/or configurations are expected to occur in this nucleus. Its transitional character has been experimentally established by the appearance of a low energy $0^{+}_{2}$ excited state~\cite{EPJA_25_111_2005} and the small value of the $E(2^{+}_{1})$ and the large reduced transition probability $B(E2,2^{+}_{1}\rightarrow0^{+}_{1})$~\cite{PLB_395_163_1997}. Very recently, the values of the monopole strength $\rho^{2}(E0,0^{+}_{2}\rightarrow0^{+}_{1})$ and the transition $B(E2,2^{+}_{1}\rightarrow0^{+}_{2})$ have been measured~\cite{PRL_105_102501_2010}. Based on shell model calculations and a simplistic two-level mixing model, these new data have been interpreted as an indication of a prolate (neutron 2p-2h) -spherical (0p-0h) shape coexistence. Additionally, some new levels have been unambiguously identified in Ref.~\cite{PRC_83_061305_2011}, specially the $4_{1}^{+}$ level which mainly decays into the $2^{+}_{1}$ state. According to shell model calculations, this state would correspond to the band head for a $K=4$ rotational band (neutron 1p-1h configuration) although the $4_{2}^{+}$ state with the neutron 2p-2h configuration of the prolate band appears almost degenerated in the theoretical predictions. \\
On the other hand, both mean field calculations with Skyrme and Relativistic interactions~\cite{PLB_333_303_1994,PRC_60_014310_1999} and beyond mean field studies with Gogny functionals~\cite{EPJA_9_35_2000,PRC_65_024304_2002} have been performed. 
Mean field results in $^{44}$S have shown the erosion of the $N=28$ shell closure and the manifestation of possible shape mixing and/or coexistence, obtaining either rather flat potential energy surfaces (PES) or prolate-oblate minima along the axial quadrupole degree of freedom. Beyond mean field calculations -without particle number restoration- using either generator coordinate method combined with axial angular momentum projection or a collective hamiltonian have confirmed the relevance of allowing for configuration mixing in this nucleus.\\
In this Rapid Communication we report the results of state-of-the-art Symmetry Conserving
Configuration Mixing (SCCM) calculations to study the structure of $^{44}$S. The method is based on the Gogny D1S interaction~\cite{NPA_428_23_1984} and contains simultaneous particle number and angular momentum projection (PNAMP) of intrinsic Hartree-Fock-Bogoliubov type states (HFB), including axial and triaxial shapes~\cite{PRC_81_064323_2010}. Variations of
our approach  implemented with Skyrme~\cite{PRC_78_024309_2008} and Relativistic Mean Field~\cite{PRC_81_044311_2010} interactions have been also reported. The HFB-type states -$|\Phi(\beta_{2},\gamma)\rangle$- are found with the variation after particle number projection method (PN-VAP)~\cite{RING_SCHUCK,NPA_696_467_2001}, being $(\beta_{2},\gamma)$ the quadrupole deformation parameters. Hence, contrary to the shell model, the collective intrinsic deformation is well established within this framework and this fact allows the description of the states in the laboratory frame in terms of their intrinsic shapes unambiguously. The Ansatz for the wave functions is provided by the generator coordinate method (GCM)~\cite{RING_SCHUCK}. In this framework, the states are assumed to be linear combinations of particle number and angular momentum projected HFB-type states:
\begin{equation}
|\Psi^{IM\sigma}\rangle=\sum_{\beta_{2},\gamma,K}g^{I\sigma}_{K}(\beta_{2},\gamma)P^{I}_{MK}P^{N}P^{Z}|\Phi(\beta_{2},\gamma)\rangle
\label{GCM_state}
\end{equation}
where $I$, $M$, $K$ are the total angular momentum and its projection on the $z$-axis in the laboratory and intrinsic frame respectively, $P^{I}_{MK}$ and $P^{N(Z)}$ the angular momentum and neutron (proton) projectors~\cite{RING_SCHUCK} and $\sigma$ labels different states obtained for a given value of $I$. The parameters $g^{I\sigma}_{K}(\beta_{2},\gamma)$
are determined by the Ritz variational principle which leads to the Hill-Wheeler equation. The resulting matrices 
are diagonalized  -one for each value of the angular momentum- providing  the energy levels and collective wave functions defined in the $(\beta_{2},\gamma)$ plane. In addition, transition probabilities $B(E2)$ and monopole strength $\rho^{2}(E0)$ can be also calculated within this framework. Detailed expressions and performance of our approach can be found in Ref.~\cite{PRC_81_064323_2010} (and references therein). This procedure upgrades in several ways the calculations performed in Refs.~\cite{EPJA_9_35_2000,PRC_65_024304_2002}. On the one hand, intrinsic HFB-type wave functions are found with the PN-VAP method instead of plain HFB improving the description of pairing correlations in the system~\cite{PLB_545_62_2002,PRL_99_062501_2007}. On the other hand, we include triaxial shapes -in Ref.~\cite{PRC_65_024304_2002} only axial configurations were allowed- and the GCM mixing is performed exactly without assuming a gaussian overlap approximation as in Ref.~\cite{EPJA_9_35_2000}. Finally, the absence of particle number restoration leads to spurious mixing of solutions with different number of particles that could affect significantly the final spectrum~\cite{PLB_2011}. Concerning technical details about the calculations, we use a regular triangular mesh in the triaxial plane including $N_{\mathrm{GCM}}=83$ HFB-type states. Each of these states is assumed to conserve both time-reversal and spatial parity symmetries and is expanded  in a single particle basis with nine major spherical harmonic oscillator shells. The number of integration points in the Euler -$(a,b,c)$- and gauge -$\varphi$- angles are chosen to ensure the convergence of both diagonal and non-diagonal projected matrix elements of the total angular momentum and particle number operators to the nominal values $I(I+1)$ and $N=28,Z=16$. In this case, these values are $N_{a}=8$, $N_{b}=N_{c}=16$ and $N_{\varphi}=9$ (see Ref.~\cite{PRC_81_064323_2010} for details). \\
\begin{figure}[t]
\begin{center}
  \includegraphics[width=\columnwidth]{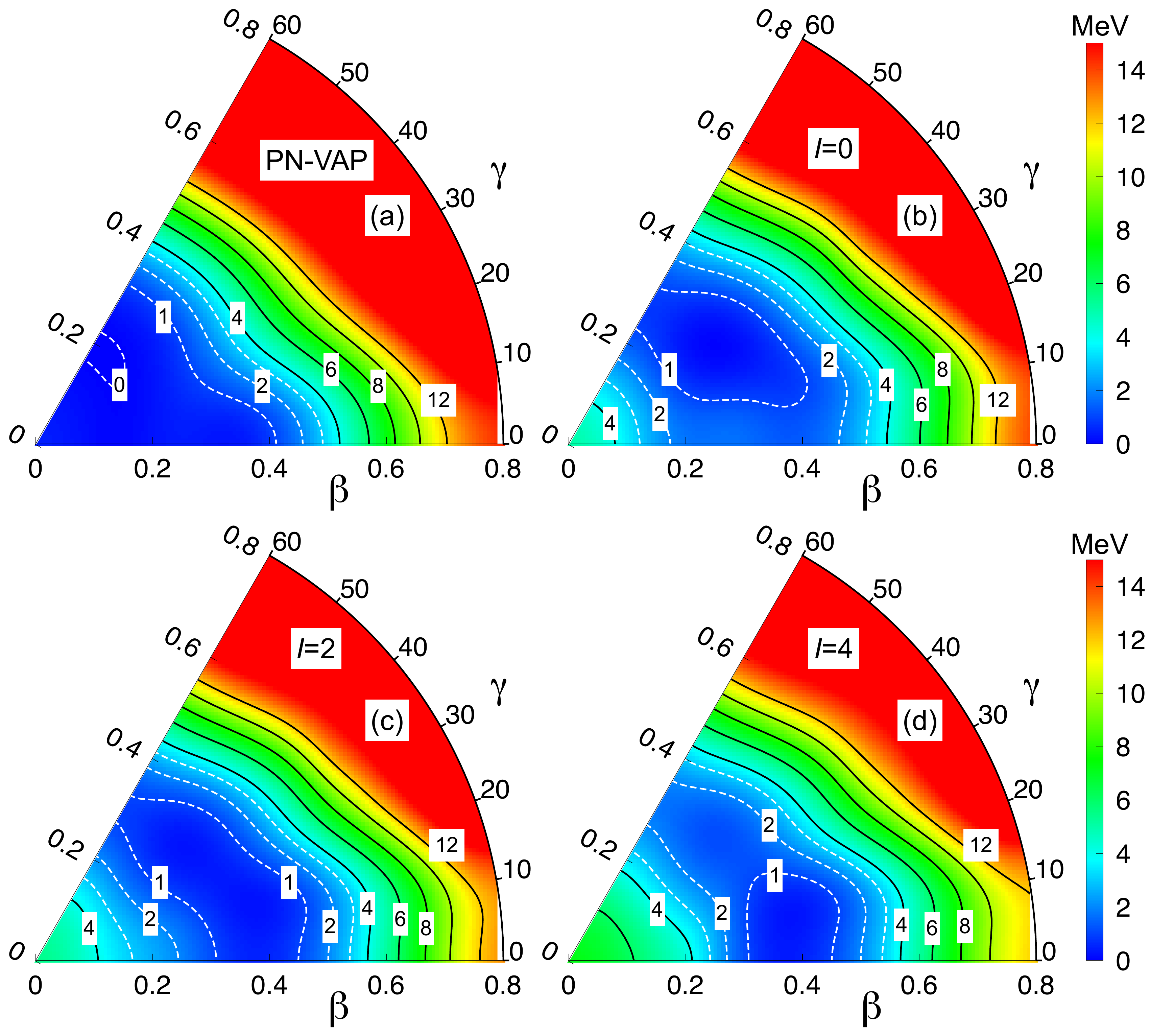}
\end{center}
\caption{(color online) Potential energy surface -normalized to the absolute minimum- in the triaxial plane calculated with (a) PN-VAP approximation and (b)-(d) PNAMP approximation with $I=0,2,4$ respectively. Contour lines are separated by 1 MeV (dashed) and 2 MeV (solid).}\label{Figure1}
\end{figure}
We first study the dependence of the energy of $^{44}$S with respect to the quadrupole deformation. In Fig.~\ref{Figure1}(a) the potential energy surface along the triaxial plane calculated in the PN-VAP approximation is shown -the equivalent  to the mean field energy in this method. We observe a very flat potential in a wide region defined by the triangle which vertices are $(\beta_{2},\gamma)\approx\lbrace(0,0^{\circ}),(0.4,0^{\circ}),(0.3,60^{\circ})\rbrace$. For larger values of $\beta_{2}$ we see a fast increase of the energy, having a slightly softer potential in the prolate part than in the oblate one. This behavior is a common feature of transitional nuclei from spherical to deformed shapes. The minimum of the surface is not located at the spherical point but at $(0.2,46^{\circ})$ which is an indication of the erosion of the $N=28$ shell closure already at this stage of the calculations. To study the effect of angular momentum restoration in the system, we calculate at each point of the $(\beta_2, \gamma)$ plane the
particle number and angular momentum projected  (PNAMP) energy.  In Fig.~\ref{Figure1}(b)-(d) we represent  for $I=0,2,4$ the corresponding PES, respectively. For $I=0$ we have only one projected PES while for $I\neq0$ $K$-mixing allows for up to $I/2+1$ different solutions. Here, the lowest energy for a given value of the angular momentum at each $(\beta_{2},\gamma)$ point is chosen. We now observe in Fig.~\ref{Figure1}(b)-(d) that the flat potential has been displaced to higher deformations excluding the configurations around the spherical shape that are now more than 4 MeV above the corresponding minima. Furthermore, we see that triaxial-oblate shapes are slightly more favored for $I=0$ around $\beta_{2}\approx0.3$, while for $I=4$ triaxial-prolate shapes are a little more preferred around $\beta\approx0.35$. For the $I=2$ case we observe an almost flat path in the whole range of $\gamma$ from $(0.4,0^{\circ})$ to $(0.3,60^{\circ})$. It is important to point out that we do not obtain PES with two (or more) well separated and almost degenerated minima, which is one of the characteristic features of shape coexistence in nuclei -like, for example, in the proton rich Se, Kr, Sr, Zr or Pb isotopes~\cite{NPA_443_39_1985,PR_215_101_1992,Nature_405_430_2000}. Strictly speaking we obtain flat potentials that suggest a wide configuration \textit{mixing} rather than a shape \textit{coexistence}.\\ 
\begin{figure*}[t]
\begin{center}
  \includegraphics[width=\textwidth]{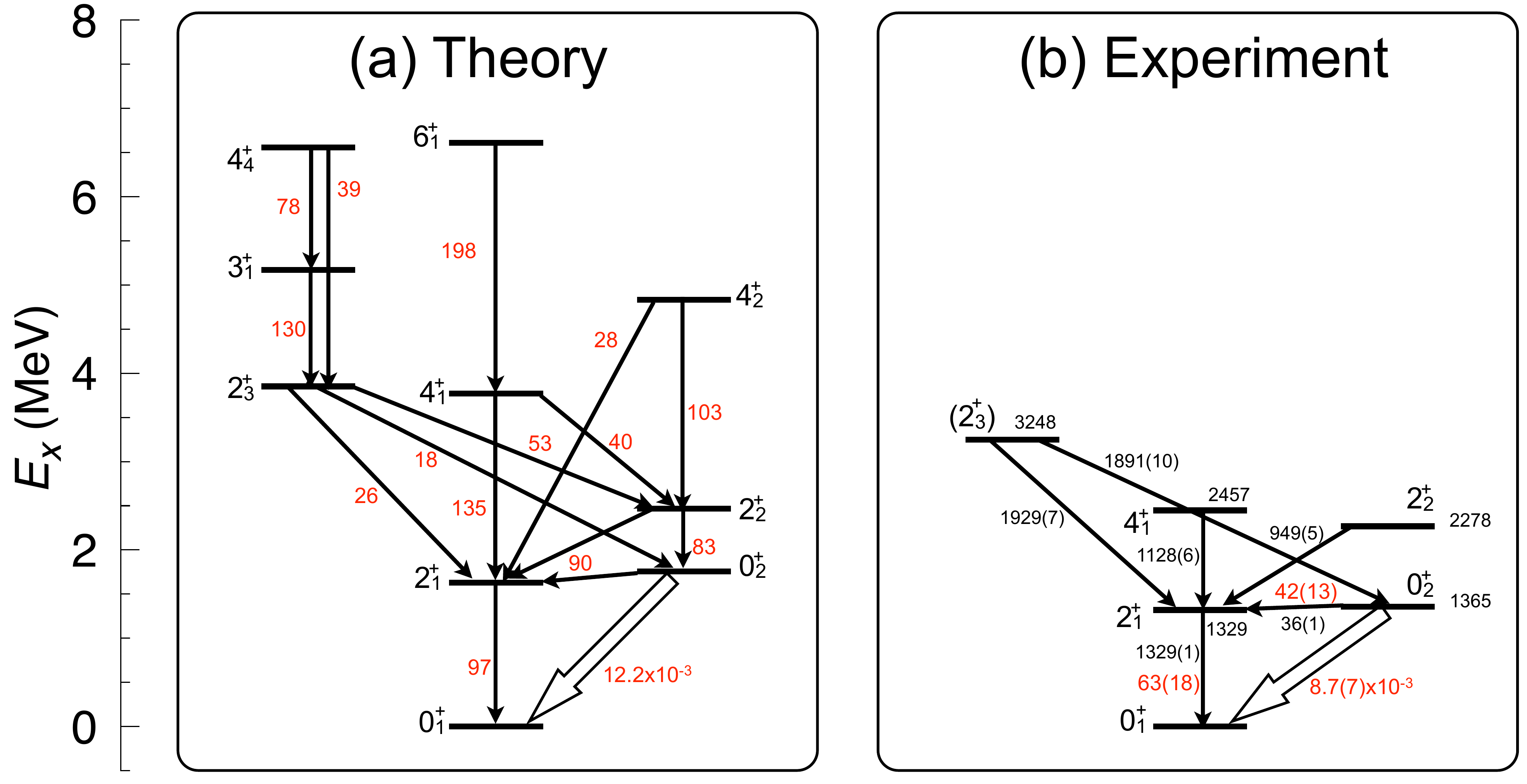}
\end{center}
\caption{(color online) Theoretical (left) and experimental~\cite{PRL_105_102501_2010,PRC_83_061305_2011} (right) spectra for $^{44}$S. $B(E2)$ values are in $e^{2}$fm$^{4}$, thick arrows represent $E0$ transition with its corresponding value for $\rho^{2}$. Experimental $\gamma$-rays and excitation energies are in keV.}\label{Figure2}
\end{figure*}
We now discuss the results of the SCCM calculations. The spectrum obtained after restoration of the symmetries and shape mixing is plotted in Fig.~\ref{Figure2}(a). Here we have sorted the levels in bands, grouping states according to their largest $B(E2)$ values. Nevertheless, we observe that there are relatively strong inter-band transitions showing that the underlying structure of these states has a pronounce mixing. The ground state is mainly populated from the first $2^{+}_{1}$ excited state with an $E2$ transition and is also connected via a monopole transition ($E0$) to the $0^{+}_{2}$ state. These two excited levels are very close in energy and they are also connected through an $E2$ transition. The first $2^{+}_{1}$ state in the calculations is connected to several levels, mainly to the
$0^{+}_{2}$, $2^{+}_{2}$ and $4^{+}_{1}$ but also to a lesser extent to $2^{+}_{3}$ and $4^{+}_{2}$. This shows again the configuration mixing present in these states. 

To compare the results of our calculations to the available experimental information, we summarize in
Fig.~\ref{Figure2}(b) the experimental level scheme of $^{44}$S including all firmly established excited
states Refs.~\cite{EPJA_25_111_2005,PRL_105_102501_2010,PRC_83_061305_2011}. The precise energy of the first excited 2$^+$ state, E(2$^+$)=1329(1) keV, has
been determined in studies of the decay of the 0$^+_2$ isomeric state at 1365 keV \cite{EPJA_25_111_2005,PRL_105_102501_2010}  in
which the 2$^+$ state decays at rest avoiding uncertainties due to the Doppler broadening of the
lineshape. In Ref.~\cite{PRC_83_061305_2011} three additional $\gamma$ rays with 949(5), 1128(6) and
1929(7) keV have been observed in coincidence with the 2$^+_1$ $\rightarrow$ 0$^+_1$ ground
state transition and assigned as directly populating the 2$^+_1$ state from levels at 2278, 2457, and
3248 keV, respectively. For the first two of these newly established states, spin values of 2$^+$ and
4$^+$ have been assigned on the basis of measured longitudinal momentum distributions
\cite{PRC_83_061305_2011}. A further $\gamma$-transition with an energy of 1891(10) keV, observed in coincidence
with $^{44}$S residues, has tentatively been placed in the level scheme to connect the 3248 keV level
to the isomeric $0^+_2$ state based on the fact that the energy difference between the 1929(7) and
1891(10) keV $\gamma$-rays ( 38(12) keV ) is similar to the one between the 0$^+_2$ and 2$^+_1$
states ( 36(2) keV ). A spin of 2$^+$ has been tentatively assigned to the 3248 keV
state due to its two decay branches to states with spin 2$^+_1$ and 0$^+_2$. Finally, in the same
work, a 2150 keV transition was tentatively  placed as populating the ground state based on its
non-observation in coincidence with the 1329 keV 2$^+_1$ $\rightarrow$ 0$^+_1$ transition.
However, this transition has been omitted in Fig. 2(b) since considering the quality of the spectra
shown in Ref.~\cite{PRC_83_061305_2011} its placement seems much less justified as compared to the case of the 1891 keV transition, a judgement which is also supported by the huge difference between the experimental cross section
for its population and the one calculated using reaction theory.
The comparison between the experimental data and our theoretical results shows a nice qualitative
agreement with a one-to-one correlation between observed and calculated states. In general the
calculated excitation energies are larger than the experimental ones which can be explained by the
lack of time-reversal symmetry breaking HFB states in the present work. Including those configurations
we expect a compression of the theoretical spectrum~\cite{PRC_76_044304_2007}. 
Starting with the two first excited states at 1329 respectively 1365 keV, we obtain slightly larger values for
both their energies as well as for the transition probabilities for their decay branches (see Fig.~\ref{Figure2}(b)).
A similar picture is found in the shell model calculation of Ref.~\cite{PRL_105_102501_2010}
-almost degenerated $2^{+}_{1}$ and $0^{+}_{2}$ states and larger calculated $B(E2)$ transition probabilities 
than experimentally obtained. However, the advantage of our approach as compared to the
shell model is that we do not need to use
effective charges and that we can also compute the value of the $\rho^{2}(E0)$ within the same
framework~\cite{PRL_103_012501_2009}. The latter quantity agrees remarkably well with experiment.\\
Turning now to the higher-lying excited states, we will discuss in the following the main differences between
the calculated and the experimental spectrum. 
For the $2^{+}_{2}$ state the calculations predict a second decay branch besides the observed
$2^{+}_{2}$ $\rightarrow$ $2^{+}_{1}$ transition, namely to the $0^{+}_{2}$ state at 1365 keV.
Such a transition, with an energy of 913 keV, has not been observed in Ref.\cite{PRC_83_061305_2011} and this
non-observation is significant given the high statistics obtained in that experiment.
The calculated 4$^+_1$ state decays mainly to the $2^{+}_{1}$ level as is observed experimentally.
In addition, a second decay branch to the $2^{+}_{2}$ state is obtained which has not been observed
experimentally. However, since experimentally the 4$^+_1$ state is observed at much lower excitation
energy, this low-energy $\gamma$-transition of 179 keV is hindered due to phase-space arguments
and furthermore may have escaped observation due to the relatively high energy thresholds used in
the experiments.\\
Also in the case of the third excited 2$^+$ state, the calculations predict one more decay branch than
identified experimentally. While the calculated decays of the 2$^+_3$ state to the 2$^+_1$ and
0$^+_2$ levels can be identified with the 1929(7) and 1891(10) keV $\gamma$-rays observed in Ref.~\cite{PRC_83_061305_2011}, a 980 keV 2$^+_3$ $\rightarrow$ 2$^+_2$ transition has not been reported. But again, the transition is hindered due to the lower transition energy as compared to the other two decay branches which may
explain its non-observation. According to our results, the $2^{+}_{3}$ state is the band head of a quasi-$\gamma$ band with
levels $3^{+}_{1}, 4^{+}_{4}$ on top it.\\

\begin{figure}[b]
\begin{center}
  \includegraphics[width=\columnwidth]{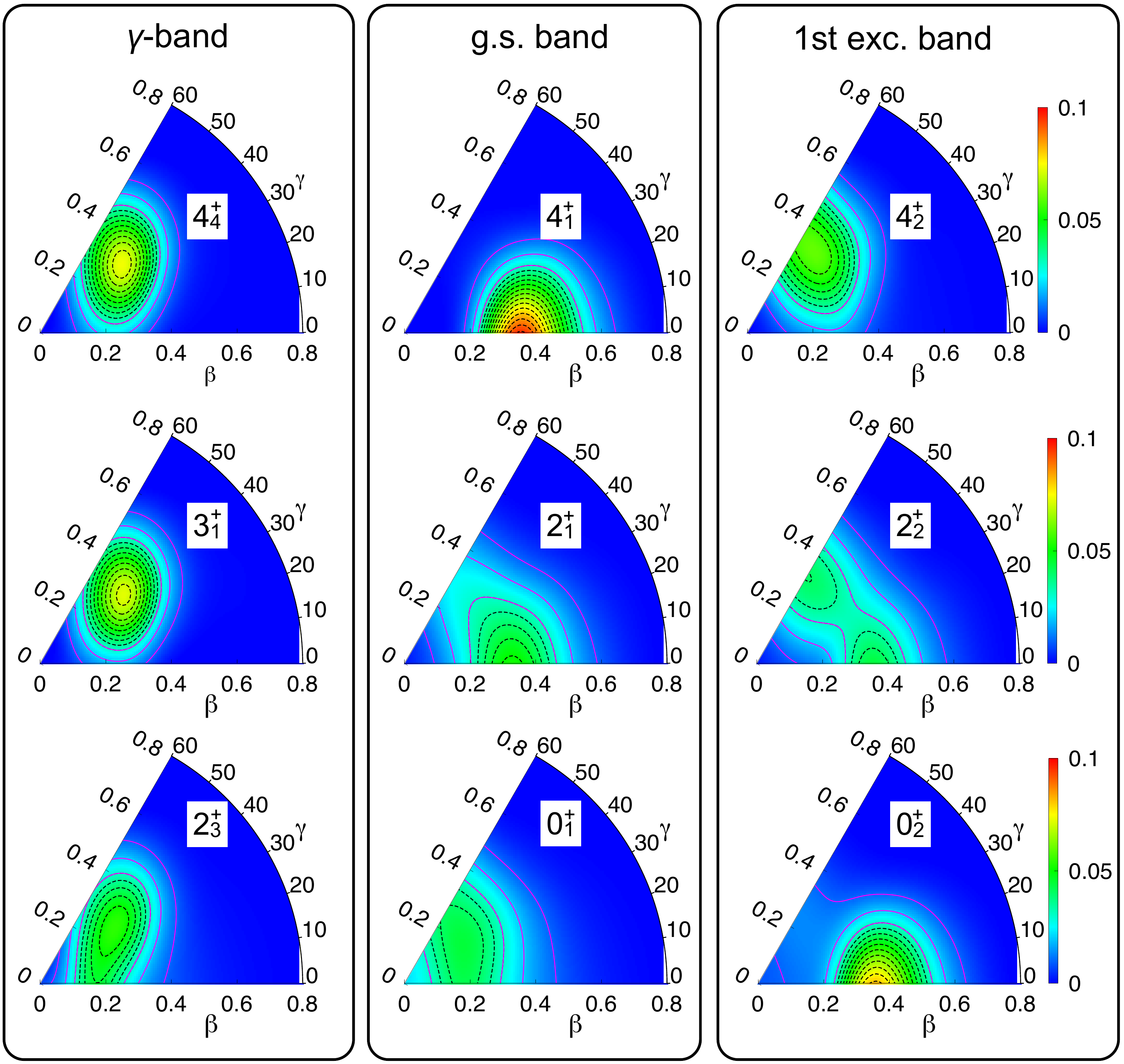}
\end{center}
\caption{(color online) Collective wave functions in the triaxial plane for the states represented in Fig.~\ref{Figure2}.}
  \label{Figure3}
\end{figure}
\begin{table}[t]
   \begin{tabular}{c|cccc} 
   \hline \hline
$I^{+}_{\sigma}$ & $K=0$ & $|K|=2$ & $|K|=4$ & $Q_{\mathrm{spec}}$ ($e$fm$^{2}$) \\
\hline
$0^{+}_{1}$ & 1.000 & -- & -- & 0 \\  
$2^{+}_{1}$ & 0.942 & 0.058 & -- & -9.2 \\ 
$4^{+}_{1}$ & 0.982 & 0.017 & 0.001 & -26.1 \\
\hline
$0^{+}_{2}$ & 1.000 & -- & -- & 0 \\
$2^{+}_{2}$ & 0.905 & 0.095 & -- & 1.2 \\
$4^{+}_{2}$ & 0.794 & 0.067 & 0.005 & 18.6 \\
\hline 
$2^{+}_{3}$ & 0.442 & 0.558 & -- & -13.0 \\
$3^{+}_{1}$ & 0.000 & 1.000 & 0.000 & -0.5 \\
$4^{+}_{4}$ & 0.185 & 0.682 & 0.133 & 14.8 \\
\hline
\hline
   \end{tabular}
   \caption{Distribution of $|K|\leq4$ and spectroscopic quadrupole moments of the states appearing in Figure~\ref{Figure3}.}
   \label{Table1}
\end{table}
To shed more light on the structure of the calculated states we represent in Fig.~\ref{Figure3} the collective wave functions of the most relevant ones. Additionally, their distribution in $K$ and spectroscopic quadrupole moments are also given in Table~\ref{Table1}. In Fig.~\ref{Figure3} we observe that the distribution of probability in the $(\beta_{2},\gamma)$ plane for each state belonging to the same band changes significantly. However, shape mixing allows the overlap between these states and the quadrupole transitions between them. Hence, the ground state $0^{+}_{1}$ is distributed almost uniformly in the whole range of $\gamma$ in a trapezoidal region between $(0.1,0^{\circ}),(0.3,0^{\circ}),(0.1,60^{\circ}),(0.35,60^{\circ})$. Therefore, the ground state for $^{44}$S does not correspond to a spherical shape -as it should be if the $N=28$ shell closure is preserved- but a mixing of deformations where the triaxial degree of freedom plays an important role. The rest of the states of this band, which have a $K=0$ configuration, evolve towards axial prolate shapes. Their maxima are found at $\beta_{2}\approx 0.35$, having the $2^{+}_{1}$ more mixing with triaxial and axial oblate configurations than the $4^{+}_{1}$. 
These prolate states appear in shell model calculations~\cite{PRL_105_102501_2010} although in that case a prolate shape is assumed for the $0^{+}_{1}$ contrary to the result found here. Furthermore, in Ref.~\cite{PRC_83_061305_2011} the $4^{+}_{1}$ state is predicted to have a neutron 1p-1h excitation configuration, being the band head of a $K=4$ band. This is not the case in the present work where the first $4^{+}$ state with the highest $K=4$ component is found at 5.4 MeV having a small $B(E2)$ value to the $2^{+}_{1}$ state.\\
The first excited band is also $K=0$ mainly. The band head $0^{+}_{2}$ state is axial prolate deformed with the maximum of the probability found at $\beta_{2}\approx 0.4$. This distribution explains the large transition probability to the $2^{+}_{1}$ state. We also observe an axial prolate-oblate mixing for the $2^{+}_{2}$ state which connects to the prolate deformed $0^{+}_{2}$, $2^{+}_{1}$ and $4^{+}_{1}$ states and also to the mainly oblate configuration of the $4^{+}_{2}$ state.  A rather small spectroscopic quadrupole moment of the $2^{+}_{2}$ state is found here and also in shell model calculations. This result was interpreted as an indication of spherical shapes for both $2^{+}_{2}$ and $0^{+}_{2}$~\cite{PRL_105_102501_2010}. However, this is not the case in the present calculations.\\
Finally, a quasi-gamma band with $K=0,2$ mixing for $I$-even and $K=2$ for $I$-odd is obtained grouping the $2^{+}_{3}$, $3^{+}_{1}$ and $4^{+}_{4}$ states. The collective wave functions of these states are rather similar, with the corresponding maxima found at $\sim(0.3,40^{\circ})$ and the probability decreasing slightly more pronouncedly in the $\gamma$ direction than in the $\beta_{2}$ one. The values for the spectroscopic quadrupole moments are typical of intrinsic oblate shapes when $K=2$ configurations are also present, i.e., negative, zero and positive for the $2^{+},3^{+},4^{+}$ states respectively. \\
To summarize, we have study the nuclear structure of $^{44}$S with a state-of-the-art SCCM method including beyond-mean-field effects such as particle number and angular momentum restoration and shape mixing of axial and triaxial configurations. Wide configuration mixing rather than shape coexistence is found in most of the low-lying states supporting the erosion of the $N=28$ shell closure and the transitional character of this isotope. A deformed and $\gamma$-soft $0^{+}_{1}$ as well as a prolate deformed $0^{+}_{2}$ states are obtained. Our results are in a very good qualitative agreement with the recently available experimental data and with the most recent shell model calculations although the interpretation of some of these data is different. Finally, we also find a quasi-$\gamma$ band with decaying properties that could be attributed to the experimental $(2^{+})$ level measured at 3248 keV.
\begin{acknowledgments} The authors acknowledge financial support from the Spanish Ministerio de Educaci\'on y Ciencia under contract  FPA2009-13377-C02-01as well as enlightening discussions with Andrea Jungclaus. T.R.R. thanks fundings from a HIC4FAIR scholarship and Victor Modamio for useful comments.
\end{acknowledgments}

\end{document}